# Bad, mad, and cooked: Moral responsibility for civilian harms in human-AI military teams

*Accountability can… be a powerful tool for motivating better practices, and consequently more reliability and trustworthy [computer] systems*
*(Nissenbaum, 1996, p. 26)*

Dr S. Kate Devitt[1]

University of Queensland, Email k.devitt@uq.edu.au

## Abstract

This chapter explores moral responsibility for civilian harms by human-artificial intelligence (AI) teams. Although militaries may have some bad apples responsible for war crimes and some mad apples unable to be responsible for their actions during a conflict, increasingly militaries may 'cook' their good apples by putting them in untenable decision-making environments through the processes of replacing human decision making with AI determinations in war making. Responsibility for civilian harm in human-AI military teams may be contested, risking operators becoming detached, being extreme moral witnesses, becoming moral crumple zones or suffering moral injury from being part of larger human-AI systems authorised by the state. Acknowledging military ethics, human factors and AI work to date as well as critical case studies, this chapter offers new mechanisms to map out conditions for moral responsibility in human-AI teams. These include: 1) new decision responsibility prompts for critical decision method in a cognitive task analysis, and 2) applying an AI workplace health and safety framework for identifying cognitive and psychological risks relevant to attributions of moral responsibility in targeting decisions. Mechanisms such as these enable militaries to design human-centred AI systems for responsible deployment.

*Index terms*: accountability, AI ethics, AI in military, autonomy, civilian harms, civilian protection, decision making, ethics, human-AI teams, human factors, international humanitarian law, law, military, military operations, military systems, moral responsibility, mosaic warfare, psychological effects, responsibility attribution, risk analysis, situational awareness, systems integration, targeting, LLMs, safety, warfare, workplace safety,

*Word count:* 11,435 words (excluding references)

---

[1] Thanks to Jan Maarten Schraagen, Alec Tattersall and Kobi Leins for their review comments that improved this chapter immeasurably

Bad, mad, and cooked: Moral responsibility for civilian harms in human-AI teams

# Introduction

> *"… as impressive and professional the Army's training and preparation processes are, and no matter how tough and skilled our frontline soldiers are, there is a limit to their mental resilience"* (Fitzgibbon, 2020).

War puts human decision makers in friction and fog; often without good choices, picking between the least bad options from a limited understanding of a complex and congested battle space and strained cognitive capacities. AI has the potential to resolve some uncertainties faster, to bring greater decision confidence. Human-AI teams, done well, could make compliance with international humanitarian law easier, to help protect civilians and protected objects in conflicts (Lewis, 2023; Lewis & Ilachinski, 2022; Roberson et al., 2022).

AI will amplify at speed and scale whatever values are prioritised and programmed into it as well as those emerging from human-AI decision teams. For example, if AI software engineers spend a disproportionate amount of their time categorising military targets and little time classifying protected objects, then civilians and objects of high humanitarian value will not be prioritised to be seen or to protect. If system engineers and decision scientists do not consider risk to civilians from the beginning of the development of new capabilities, then the risk of causing unintentional human harms is increased. AI has limited reasoning and ability to respond to surprise, as demonstrated by protesters placing traffic cones on the hood of Waymo driverless cars in San Francisco rendering them immobile (Kerr, 2023). Networked autonomous assets set to fire and the humans set to oversee the application of force, may see only the military objectives, and fail to see the proportionate effect on humanity and culture within the domain of attack.

This chapter argues that we risk putting humans into battlefields where their ability to act ethically, according to their own judgments of the correct course of action, is problematically reduced due to the integration of information and decision layers that nudge or coerce them into compliance with an algorithmic assessment. The chapter asserts that society is likely to punish human operators after incidents in its desire to ensure accountability. The risk is that there is little accountability for those who create, deploy and maintain war fighting decision systems. in the chapter these human operators are 'cooked apples' and they are at risk of a range of psychological effects from disengagement from their role to moral injury. We must first acknowledge the risk of 'cooked apples' and seek to ensure mechanisms for accountability of decision systems are employed by militaries.

The chapter will first focus on the sort of AI likely to affect human decision-making and attributions of responsibility, so AI that takes over typically cognitive tasks of humans in a warfighting context. The chapter will provide some definitions on responsibility and AI. Then some background regarding responsibility in war will be provided and I will introduce the concepts of 'bad', 'mad' apples and 'cooked apples'.  Before delving into how AI can contribute to cooked apples, I introduce ways in which AI can reduce civilian harm—indicating that the chapter is not a whole-hearted critique of AI, only that its implementation and operation must be scrutinised. The chapter goes on to discuss the challenges of accountability in the history of computers and the specific risk of Large Language Model (LLM) AI models before suggesting ways to measure and anticipate the human experience of working with AI. I suggest modifications to critical decision method in cognitive task analysis to include Decision Responsibility Probes as one means to explore moral responsibility attributions. The chapter finishes with a tool to assess AI workplace risks relevant to military contexts of use to indicate how a risk-based approach could contribute to measures of an individual's engagement and thus responsibility for targeting decisions as well as when they feel overly responsible for technologies that are a part of a complex system. Combining tools from human factors and workplace health and safety will enable militaries to track, measure and take responsibility for ensuring personnel remain 'good' and not 'cooked'.

## Definitions

### Responsibility

The *Edinburgh Declaration on Responsibility for Responsible AI* (Vallor, 2023, 14 July) differentiates between five types of responsibility for AI and autonomous systems.

1. Causal responsibility: What event made this other event happen
2. Moral responsibility: Who is accountable for answerable for this?
3. Legal responsibility: Who is, or will be, liable for this?
4. Role responsibility: Whose duty was it (or is it) to do something about this?
5. Virtue responsibility How trustworthy is this person or organisation

The Declaration argues that responsibility is best understood as relational rather than an agent or system property. So, rather than saying an isolated agent is responsible, a better question is to ask about the context of responsibility and to consider who an agent is responsible to and how that responsibility manifests within the context of decision making.

Responsibility in this way means "articulating the ongoing, evolving duties of care that publics rightly expect and that people, organisations and institutions must fulfil to protect the specific relationships in which they and AI/AS are embedded" (Vallor, 2023).

### Moral responsibility

This paper grounds the concepts of moral responsibility within the philosophical and legal tradition that encompasses liability, blame and accountability. A person is morally blameworthy if their actions cause the harm or constituted a significant causal factor in bringing about the harm and their actions were faulty (Feinberg, 1985). Faulty actions can be intentionally or unintentionally faulty. A person can be held liable for unintentionally faulty actions if an incident occurs due to recklessness or negligence. A reckless action is one where a person can foresee harm but does nothing to prevent it. A negligent action is one where a person does not consider probable harmful consequences.

Morally blameworthy individuals must have situational awareness (they must know the situation within which they are making decisions), they must have capability to intervene to make the right moral action and they must have agency and autonomy over their decision. Thus, an individual who is coerced, forced, restrained or unable to act morally, cannot be held morally accountable for unethical acts. An individual who has been poorly informed, misinformed, partially informed or uninformed cannot be morally blameworthy unless, by their actions, they prevented themselves from being relevantly informed. Finally, considerate of the findings of the Edinburgh Declaration (Vallor, 2023), any attributions of moral blameworthiness must be situated within the duties and expectations organisation and institutions. This accords with the central premise of the chapter that the moral blameworthiness of decisions of individuals must be considered within the broader decision environments within which they act.

## Artificial intelligence

The chapter draws on the *OECD taxonomy of AI systems* (OECD, 2022) reconsidered for military AI as described in the *Responsible AI in Defence (RAID) Tool*kit (Trusted Autonomous Systems, 2023) to define an AI system. In this way a military AI system can be described by moving through the questions in the *RAID Toolkit D. Checklist Complete*, p.9-22:

   A. AI: What is the AI capability and how does the AI component function? Explain what the AI capability comprises, and how the specific AI element will function.
   B. Development inputs: What is the composition of the AI functionality. Explain the complexity and structure of the algorithms, and interlinked hardware that constitute the AI functionality.
   C. Human Machine Interaction: How does the AI capability (and more specifically the AI functionality) interact with the human operators across the spectrum of human involvement.
   D. AI Use Inputs: Identify the foreseeable inputs for the AI capability to operate when in use. Input is the data that is required for the AI functionality to operate.
   E. AI Use Outputs: Identify the foreseeable outputs from the AI capability. Output is the data resulting from the execution of the AI functionality, or the ways that people or things receive the data resulting from the execution of the AI functionality (in whatever form the data is represented).
   F. Object of AI Action: Ascertain on what or whom the impact of the AI action will be. It requires consideration of the external actors that will be influenced or affected by the AI's actions within the environment of its anticipated use case.
   G. Use case environment: Describe the military context in which the AI will be employed.
   H. System of control: control measures, system integration and AI frameworks. Explain how the AI capability fits within the broader system of control applied to military operations and activities.

The AI I consider in this chapter is not just software engineering, algorithms, models, data or use of specific training methods such as machine learning, but the systemic replacement of cognitive functions that humans would normally be tasked with in targeting decisions to be morally and legally responsible for those decisions. Because of this, the types of AI not considered in this paper include the AI unrelated to targeting, such as that used in the functions of autonomous systems to plan, navigate, self-monitor or self-repair. It is envisioned that AI in a targeting role, by potentially fusing multiple information sources and types of information, some beyond human perceptual faculties (e.g. hyperspectral imagery, IR), may be able perform targeting assessments with greater fidelity and reliability than humans can achieve which motivates their use.

> *Perceiving*. This is AI that brings situational awareness and clarity. Examples include the classification of simple objects such as a road, chair, building, bridge, traffic light or person. A human being might need to verify the classified object or need to trust that the AI has the classification correct. This is particularly true for military objects because databases of military objects have significantly less data to draw on for model training than databases for regular objects in the real world such as dogs and

> cats. If the AI misperceives an object, then the human may have additional cognitive load and responsibilities in determining when a misperception by the AI has occurred.
>
> *Remembering*. This is AI that has historical information on prior cases upon which to constrain hypotheses for decision-making. This includes the rules, processes and procedures including rules of engagement and international humanitarian law. For instance, a LLM AI trained on the corpus of documents within a military is asked questions about precedents. A human expert (or multiple experts) would need to verify that the AI has interpreted the relevant documents correctly or reasonably and then vouch for them.
>
> *Understanding*. This is AI that integrates information from multiple sources into a probable scenario including causal hypotheses. For instance, interpreting thermal imagery, location, historical movements, and purpose into a theory of who is traversing a particular location and for what purpose. Examples of this include the integration of data sources into the classification of a complex scenario or situation, such as a suggesting the identity and purpose of a group of persons in a meeting at a particular location. A human would need to know the edge cases where information integration is likely to be unreliable and the most reliable circumstances when information integration is mostly likely to provide a suitable explanation of the situation.
>
> *Imagining*. This is AI that creates new concepts from prompts and parameters. For instance, a LLM AI trained on the corpus of documents within a military is asked to create innovative tactics to achieve a military objective given constraints of personnel, equipment, timeframe, conditions, and rules of engagement. A human would need to seize upon the creative output of the AI and consider its value against their own knowledge and experience, military objectives, their own goals, values and ethical assessments.
>
> *Reasoning*. This is AI that is able to apply higher order considerations in the interpretation of multiple hypotheses as well as request more information to fill in data gaps, to recommend caution and is aware of the requirements on human beings in the manner in which information is used to make decisions. For instance, a LLM AI built in the style of AutoGPT that can be tasked with a goal and create its own sub goals and projects to break the task into distinct components including a specific set of actions to question its own assumptions (Ortiz, 2023; Wiggers, 2023). A human would need to audit the decision steps and reasoning of the AI to vouch for it.

Systematic influences are solidified, codified, embedded and amplified through the introduction of AI and autonomous systems into military decision-making and action, and unlikely to be scrutinised or engaged with critically by operators in the 'fog of war'. Thus, militaries must be clear on their objectives and design frameworks for AI and clear on where the AI is helpful to help with human analytical skills to ensure decision processes and outcomes lead to military and political success. Lessons from the medical field on where AI assists and where it makes outcomes worse may be abstracted and applied to the miliary domain. For example, during covid 19 an AI model was trained using data that included patients who were scanned standing up and lying down. The lying down patients were more likely to be seriously ill, so the AI learned wrongly to predict serious covid risk from a person's position (Heaven, 2021). The point is if the human can't critical engage (which is difficult in a battlefield situation) in AI outputs, then human-in, human-on or human-constraining-the-loop are not sufficiently informed to be responsible. Even knowing how to define or evaluate 'the loop' is challenging (Leins & Kaspersen, 2021) . A similar point has been made about the dangers of people using LLM such as ChatGPT to answer queries in domains for which they lack sufficient expertise to evaluate outputs (Oviedo-Trespalacios et al., 2023).

## Responsibility in War

> *[Commanders need the] courage to accept responsibility either before the tribunal of some outside power or before the court of one's own conscience" (Clausewitz et al., 1976)*

Nations are responsible for decisions to go to war (*jus ad bellum*) in a way that individual soldiers cannot be held accountable for, whereas soldiers are responsible for their conduct within war (*jus in bello*) (Walzer, 2015). This chapter takes the importance of taking responsibility for one's own actions as a given, as a premise. The question then becomes: how is responsibility for decisions determined when the actions of war and decision making become dominated by cognitive AI and complex systems, adding to human experience and awareness already mediated by information layers? The worry is if humans in warfare begin to eschew responsibility, then the connective tissue of war as a political act begins to fray. Thus, responsibility must be retained.

Miller (2020) argues that commanders must retain meaningful human control (MHC) of weapons if AI systems begin to select and engage targets autonomously, a task previously done by humans, in order to maintain commitments to international humanitarian law (IHL). He says, "commanders will remain obligated to take necessary and reasonable measures to prevent and suppress violations of IHL by their forces. Therefore, MHC should be defined as the control necessary for commanders to satisfy this obligation" (p.545). A responsible commander must use weapons that they understand including their purpose, capabilities, and limitations of the system. The level of direct control required will depend on the context of operations within which they are used. Commanders must apply geographic and temporal constraints to uphold distinction and proportionality. In more complex and civilian saturated environments, Miller argues meaningful human control may require commanders to apply additional control measures or human supervision. There is academic debate regarding the what 'meaningful human control' must consist of, but from a legal perspective, command responsibility has a set of clear requirements that human-AI teams must comply with (Liivoja et al., 2022).

### Moral responsibility for civilian harms

In this section the role of the war making institution with regards to civilian harm is unpacked because AI for targeting will instantiate and amplify system organisational targeting norms, values, and processes.  The principle of civilian immunity applies to the deliberate targeting of civilians and civilian objects. However, civilians may legally be harmed if the harm is unintentional and incidental to the military's objectives. Collateral damage is often seen as a necessary and ordinary consequence of war. However, Neta

Bad, mad, and cooked: Moral responsibility for civilian harms in human-AI teams

Crawford (2013) pushes back against the fatalism of this premise. She distinguishes three kinds of collateral damage:

CD1) genuine accidents,
CD2) systemic collateral damage and
CD3) the foreseeable if unintended consequence of rules of engagement, weapons choices, and tactics; and double effect/proportionality killing accepted as military necessity.

She claims that systemic collateral damage has become a moral blind spot, miscategorised as accidental, or treated as natural and unaffected by policy choices. She also argues that proportionality/double effect (Quinn, 1989) collateral damage also occurs in a moral blind spot, because of the wide legal latitude to harm civilians afforded by prioritising 'military necessity'. Her argument is that CD2 & CD3 are produced both by the expansive and permissive conceptions of military necessity and by the organisation of war making. Systemic collateral damage is decided at the organisational and command level and stem from institutionalised rules, procedures, training, and stresses of war.

Her point is that there are great gains to be made by altering these systematic influences on collateral damage. Recent changes to the Department of Defence Handbook requiring a 'presumption of civilian nature until proven' military (Leins & Durham, 2023) also create a dependence on data sources and automation that make this complexity incredibly real – from the data sets and sources used, to their security, as well as their analysis and any AI or automation, all of these raise questions regarding reliance and 'governance by data' or by assumption.

Crawford notes that typically militaries explain incidents where civilians are harmed disproportionately as being the result of individuals, not processes. So, blameworthiness is laid upon a 'few bad apples' or in some cases, even, 'mad apples'. The bad apples are those who intentionally and deliberately kill civilians. The mad apples have lost their ability to navigate decisions in war. Her point however is that "to blame individual soldiers for snapping is to be, in a sense, blind to how the moral agency of soldiers is shaped and compromised by the institutions of war making" (p.466). Unfortunately, because collateral damage is legal under international criminal law, with its focus on individual moral and legal responsibility or intentional acts, the bar is set very high for proving deliberate intention.

However, Crawford points out that it is "not only the 'reality of war' but also the structure of the law with regards to non-combatant immunity and military necessity that creates the potential for large-scale, regular, systematic collateral damage in war as the 'ordinary' consequences of military operations" (p.466). She locates CD2 in

1. Rules of engagement
2. Standard operating procedures
3. International humanitarian law

Crawford's ambition is to extend moral responsibility for decisions beyond individuals and into military organizations. She notes that the way militaries are organised both enables and constrains individual moral agency. She also claims that military organisations should themselves be moral agents. Such a claim has arguments for and against. Some in the 'against' camp say that an organisation cannot be a moral agent because it cannot feel guilt, shame and so forth, unlike the individuals who comprise it (Orts & Smith, 2017). On the other hand, individuals do seem to blame corporations and organisations for decisions enabled by the entity that cause moral harms. For example, in this chapter I take Crawford's supposition as a premise. I treat both individuals and military organisations as bearing moral responsibility. In doing so I acknowledge that there are philosophical concerns with such a premise.

Good apples follow orders but may systemically cause avoidable harms to civilians due to the systems within they are making decisions.

> AI/CD1 AI-enabled systems can increase or decrease accidental collateral damage by autonomous minimising harms and offering greater situational awareness for human oversight and intervention.
>
> AI/CD2 AI can systematically increase or decrease collateral damage if it is hosted within decision systems that reduce the activation of higher order reasoning regarding targeting decisions or reduce other psychological barriers to targeting.
>
> AI/CD3 Parameters coded within AI can increase or decrease the acceptability of double effect/proportionality killing as military necessity.

In this chapter I will consider how AI systems can exacerbate CD1, CD2 & CD3 by otherwise good apples. I will briefly touch on how AI systems can abet, reveal, or hide bad apples. But, also, note how easily incidents judged to be accidents are really the result of systemic failures, thus miscategorised.

The consequences I analyse are both whether individuals are unfairly blamed for actions for which they should not be held morally accountable for; and whether individuals themselves blame themselves and suffer moral injury for their role within systems that afforded them little recourse to avoid the scale of civilian harms that may occur. Individuals suffering misattributed accountability and/or moral injury, I refer to as 'cooked apples'. Military personnel ought to be on the lookout for environments that 'cook' them and to blow the whistle on systems where harms are magnified by these technologies.

## AI to reduce civilian harm

A premise of this chapter is that good apples intend to prevent civilian harm and conduct their professional duties in accordance with rules of engagement, the laws of armed combat, international humanitarian law and their own moral compass. Therefore, good apples would wish to reduce harm to civilians and be interested in improving decision making so that this was achieved—including considering using AI for that purpose. The Centre for Naval Analysis (Lewis, 2023) analysed over 2,000 real world incidents of civilian harm and classified them into twelve pathways divided into misidentification and collateral damage. From their analysis they identified four applications of AI in the role of *perceiving*, *remembering,* and *understanding* that could reduce civilian harms and potentially moral injury, empowering personnel to make more informed decisions without adding to their cognitive load:

1. *Alerting the presence of transient civilians*. Many civilians are harmed because intelligence is unable to keep up with the movements of civilians. Better monitoring of persons around the target area would bring them to the attention of operating forces that can fixate on a target rather than protected persons in the area.

2. *Detecting a change from collateral damage estimate*. Collateral damage estimates can be wrong. AI can be used to identify the difference between earlier estimates and new images closer to the time of an attack that might indicate a change in who is within the area.
3. *Alerting a potential miscorrelation*. Militaries can have correct information about a threat, but then misidentify the location of that threat (e.g. a vehicle with legitimate targets within may be swapped out with civilians). Better AI-enabled surveillance could help identify that a miscorrelation has taken place.
4. *Recognition of protected symbols*. Modern battlefield sensors are attuned to the infrared spectrum because it allows the confident identification of machine signatures such as engines as well as humans in varying light conditions and the apparatus does not itself (thermographic camera) have a signature like radar. AI methods could identify protected symbols for designating protected objects alerting the operator or chain of command.

Additionally investing in a trusted communications network for human-AI teams would enhance the potential to protect protected persons and objects (Devitt et al., 2023). The opportunity of AI to reduce civilian harm must be kept in mind as the chapter considers the way that teaming with AI might reduce human responsibility and accountability as well as causing harms to personnel. To begin this investigation, I will consider that computers have long presented a challenge to accountability mechanisms.

## Computers, Autonomy and Accountability

Concerns about the effect of computers on accountability have been identified for decades as computational devices have become ubiquitous. In 1966, Josef Weizenbaum referred to the 'magical thinking' around basic automated systems, and the challenge of humans often being uncritical of their outputs (Weizenbaum, 1966). In 1996, Nissenbaum argued that computers contribute to obscuring lines of accountability stemming from both the facts about computing itself and the situations in which computers are used. She identified four barriers to accountability: 1) the problem of many hands, 2) the problem of bugs, 3) blaming the computer, and 4) software ownership without liability (Nissenbaum, 1996). Computational technologies create systemic accountability gaps. Note the similarities of concern expressed by Crawford (2013), Parsons & Wilson (2020) and Nissenbaum. All three identify the *systems* of decision making as problematic and requiring a need to be addressed. The Uber self-driving car example for its recency and the lessons we might take for the future of soldiers increasingly serving with autonomous systems.

### The Uber self-driving car incident

In the Uber case, a pedestrian, Elaine Herzberg, walking her bicycle across a highway in Arizona, was killed because, first, the experimental autonomous driving system failed to recognise her as a pedestrian (confused, the AI kept reclassifying the moving object as separate cars exiting the highway rather than a continuous entity); second, the safety parameters were turned down because they were deemed as overly sensitive and distracting to the test drivers (causing a 'boy who cried wolf' problem where test drivers stopped believing that there was a genuine risk when the alerts sounded) and finally, the test driver was on her mobile phone when the incident occurred and therefore did not see the pedestrian herself in order to try and allay the accident.

Bad, mad, and cooked: Moral responsibility for civilian harms in human-AI teams

The AI in the car finally identified the object as a bicycle 2.6 seconds from hitting the object, but then switched it back to classifying it as 'other' at 1.5 seconds before impact. At this time the system generated a plan to steer around the unknown object but decided that it couldn't. At 0.2 seconds to impact the car let out a sound to alert the human operator. At two-hundredths of a second before impact the operator grabbed the steering wheel which took the car out of autonomy and into manual mode (Smiley, 2022). When the legal case was finalised, there was no liability against Uber or any of those responsible for decisions regarding the perceptive or safety systems of the autonomous vehicle. The judge determined that the solo human 'operator' was solely liable for the incident, and she was entirely blameworthy (Ormsby, 2019). Was this fair attribution of responsibility. Let us consider the systems in play:

In the years before the accident, federal regulators were standing back allowing companies to voluntarily report their safety practices and recommended that states, such as Arizona, did the same (Smiley, 2022). So, Uber was able to make their own safety policies.

A year before the incident, in 2017, Uber had changed their safety practices from requiring two humans in each test car (one driver and one to look out for and discuss hazards and to write up issues for the company to review), to only having one human in each self-driving car. This policy change had several impacts. Now, a single operator had to manage the complexities of each test drive. Additionally, there was no longer a conversational partner to share situational awareness with and to prevent automation complacency. The Uber AI's self-driving competence had increased sufficiently by this time that the cars made mistakes much more rarely. Being alone for hundreds of miles increased the allure of the single operator's mobile phone and increased the likelihood of attentional safety transgressions.

In this instance, the test driver became the locus of blame for the incident even though the technical factors contributing to the incident still required investigation and intervention. Still, from a legal perspective, and consistent with the definition provided regarding being informed, if the test driver had been watching the road rather than being on their phone, then they could have seen the pedestrian and taken control of the vehicle and averted the accident. The legal manoeuvre reveals that the driver was 'cooked', the human driver was expected to behave in a way that was difficult and unrealistic for a reasonable human to do—and over 70 years of human factors research on automation bias would support this (Endsley, 2017; Hoff & Bashir, 2015; Lee & See, 2004; Merritt et al., 2013). Human attention tends to wander when they are not actively engaged in a task. When systems operate as expected, humans become complacent. The difficulty of sustaining attention for monitoring purposes over longer periods of time is also known as vigilance decrement (Martínez-Pérez et al., 2023).

What is egregious about allowing accountability to fall only an inattentive human operator is that organisations should know better. Attention wandering was first noted in British naval radar operators who increasingly missed critical radar signals (enemy combatants) as their watch periods progressed. This phenomenon has been studied scientifically since the end of World War II (Cummings et al., 2016; Mackworth, 1948; Mackworth, 1950; Thomson et al., 2015) should be anticipated and its effects mitigated.

## Risks of large language models

Large language models (LLMs) trained on vast data sets that predict sentences based on natural language inputs. While in existence for a number of years, have burst into public

Bad, mad, and cooked: Moral responsibility for civilian harms in human-AI teams consciousness in 2022 and current iterations such as 2023's GPT-4 with one trillion parameters exhibit a remarkable degree of higher order reasoning and complex thinking compared with other forms of AI to date (Bastian, 2023; Hardy et al., 2023; Ott et al., 2023). LLMs will increasingly be used to improve strategic and operational effectiveness with better data, models and scenarios and are likely to affect responsibility attributions.

To gain the trust of users, the interface of LLMs are designed around the cognitive architecture of human users using natural language explanations rather than statistical or didactic responses. Such approaches are promising because information is clear and the logic more transparent for human users. LLM integration with robots may also enable better ability for robots to choose actions that comply human intent and are relevant to solving the task (Rana et al., 2023). The complexity and seamless presentation of information to operators and the ease of LLM query response suggests a step-change in human-machine integration.

However, the ease of legibility belies the limits of the underlying models for recommending courses of action including partial causal variables, inductive error, limits in data, calibration errors, context insensitivity, simplification, and extrinsic uncertainties. Also worryingly, the better the cognitive fit of a tool, the more influential outputs can be on user acceptance (Giboney et al., 2015) regardless of system limitations. Cognitive fit refers to where information is presented in a way most easily processed by a cognitive agent. For example intelligence briefs prepared with more pictures can be easier for some to process than wordy text (Barnes et al., 2022). Paradoxically, the ease of interaction between chat LLMs and humans may reduce the engagement of higher order functions that enable humans to reflect and question information provided. Thus, LLMs appear to be interacting with end users in a rational dialectic, but in fact are using rational persuasion as a form of paternalism (Tsai, 2014) creating a particularly sophisticated and nuanced form of automation bias. LLM outputs may intrude on the users' deliberative activities in ways that devalue her reflective decision-making processes and keeps decision making concordant with system level values.

So, a danger of designing AI tools that mimic higher order human reasoning and produce outputs aligned to the cognitive preferences of decision makers is that users may be manipulated to assent. When an LLM/AI agent rationally persuades a user, it offers reasons, evidence, or arguments. It is possible to construct an AI to rationally persuade a human operator to choose a right action, yet the information represented is paternalistic or disrespectful by being incomplete, simplified or obfuscating. Therefore, Systems that appear very confident, authoritative, and omniscient need to self-identify their own limits to human operators and modulate their confidence in their answers and humans need to be trained sufficiently to know when and how to be skeptical of AI reports. LLMs that use rational persuasion are likely to be attractive to militaries to achieve cohesion and conformity of response and to offset human decision maker's limited capacity to gather, weigh or evaluate evidence, as well as for efficiency reasons—where human hesitation, cogitation, unreliable reactions will slow decision making down in high tempo conflicts.

AIs deployed with humans need to gain and maintain trust and avoid paternalism. How should this be achieved? AI systems use vastly more data and operations on that data including data integration than a single decision maker can understand, rendering individual reflective cognition problematic. When action recommendations align with expert human intuition, there may be little cause for concern. However, when recommendations diverge from expert human intuition, then humans must either trust the system—and follow its

dictates —without necessarily knowing why they are agreeing; or reject the system risking a suboptimal alternative. AI system makers may try to improve trust with honest articulation of how decisions are generated, but it is likely that information will necessarily be simplified and manipulated to facilitate consent. And, while users benefit from gaining more decision control, explanations can sometimes increase cognitive load without assisting their decision making (Westphal et al., 2023).

The fast integration of LLM as chat agents that appear to exhibit cognitive functions of thinking, understanding, correcting themselves, etc… as well as robotic systems able to autonomously adjust their actions to changing circumstances, increases the likelihood that humans working with machines will attribute agency to the AIs driving these systems or even considering them moral agents; potentially avoiding responsibility for decisions made by the AI when things go wrong and taking more credit for outcomes when things go right.

Johnson (2023) argues that human-like human-robotic interfaces make people feel less responsible for the success or failures of tasks and use AI agents as scapegoats when bad outcomes occur. He points out that use of AI will require more (rather than less) contributions and oversight from the human operator to mitigate the contingencies that fall outside of an algorithm's training parameters or fail in some way. Given the risks of LLMs for responsibility attributions, it is worth reviewing how a military LLM are being marketed.

## Palantir's Artificial Intelligence Platform

Palantir's Artificial Intelligence Platform [AIP] for Defense (Palantir, 2023) has been marketed in the context of future military decision making. Palantir emphasises that "LLMs and algorithms must be controlled in this highly regulated and sensitive context to ensure that they are used in a legal and ethical way [0:18-0:25]" and claim that their AIP has:

> "industry leading guardrails to control, govern and trust in the AI. As operators and AI take action in platform, AIP generates a secure digital record of operations. These capabilities are crucial for mitigating significant legal, regulatory and ethical risks posed by LLMs and AI in sensitive and classified settings. [0:57-1:17]"

In the scenario demonstration, Palantir notes that using their system effectively requires deep understanding of "military doctrine, logistics and battle dynamics [5:51-5:55]". Implicit is the notion that doctrine contains the legal and ethical values required to ensure human responsibility over decisions. Human agency is recorded through the process of human interrogation of the system described by Palantir as "reasoning through different scenarios and courses of action safely and at scale [6:28-6:32]". However, the video and associated documentation do not provide transparent guidance for users or decision-makers on how operators can verify or validate the data they are receiving from the LLM. The 'content protection' tab offers obscure acronyms and a 'validation' option [7:23], but without any explanation—at least from the publicly available materials. To trust these systems, acquirers of these technologies would ensure appropriate test and evaluation, verification, and validation of these systems, to be assured that they are fit for purpose under the anticipated range of uses within a defined context of operations. They would also need a tailored curriculum and training program for operators to ensure that operators knew the parameters of the system and the bounds of their own role in decisions being made in the battlespace. Instead, the Palantir marketing material focuses on their claims of trustworthiness, auditability of human reasoning, LLM outputs and decision-making, and ultimately putting responsibility for decisions back on humans using the technology. Are militaries prepared to manage LLM integration ethically, so that individuals are capable of holding moral responsibility for decisions made with LLMs?

So far the risks of LLMs discussed include soldiers making decisions without adequate ability to engage their higher order critical functions and possibly also making soldiers feel less responsible from their decisions. I will now consider how AI can make operators feel even more responsible—particularly when things go wrong.

## Moral Injury

Moral injury refers to the psychological effect on soldiers who feel ethically compromised through their professional conduct in warfare. Human-AI teams change the decision-making environment and will change the risk calculus of moral injury. This chapter takes as a premise that militaries ought to reduce the conditions of moral injury. This premise is controversial. Some in society, such as absolute pacifists would expect soldiers to always suffer moral injury in a conflict because they believe that the act of killing is unethical under any circumstances. This chapter assumes a just war perspective that acts of harm by a state are, in some cases, justified. In those cases, the military ought to ensure soldiers are fully cognisant of why war is being waged and why it believes that the conduct expected of soldiers is justified. A soldier's individual sense of ethical conduct is constructed within these institutional conditions.

Moral injury can be caused by ethical discordance between the individual regarding whether entering a conflict itself is just (*jus ad bellum*), how soldiers judge their own conduct during a conflict, but also how soldiers perceive the ethical conduct of the military they serve during a conflict (*jus in bello*). This chapter will focus on moral injury *jus in bello* considered broadly to not just include the individual responsibilities of soldiers, but also the strategic and organisational war-waging responsibilities of the military organisation. Parsons and Wilson (2020) divide these into responsibility for aligning war aims with means of war, achieving organisational capacity to achieve war aims at the least cost in lives and resources and the maintenance of legitimacy (see Box 1).

> **Box 1. Three war-waging responsibilities in *jus in bello's* strategic dimension**
>
> (1) Achieve and sustain coherence: war aims must be aligned with means as well as strategies, policies, and campaigns in order to increase the probability of achieving the aims set.
>
> (2) Generate and sustain organizational capacity: initial aims and decisions must be translated into actions that achieve the war aims at the least cost in lives and resources and the least risk to the innocent and one's political community. These decisions and actions must adapt to changing conditions as the war unfolds and bring the war to a successful end.
>
> (3) Maintain legitimacy: war must not only be initiated for the right reasons and observe the laws of war; additionally, public support must be sustained, and the proper integration of military and civilian leadership must be ensured. Executing these responsibilities sufficiently well is the second way political leaders exercise their responsibilities to their soldiers and their nation as well as the innocent put at risk by war

Bad, mad, and cooked: Moral responsibility for civilian harms in human-AI teams

> (Parsons & Wilson, 2020)

Soldiers may be critical of the way their militaries are conducting war including how they are teamed with technologies. For example negative psychological effects may occur for remote pilots operating physically alone, rather than working side-by-side with a team. Militaries might have personnel working unnatural and jarring military shifts only to switch to a civilian one as they move in-between their family homes to secure remote military locations each day (Enemark, 2019). Moral injury may stem from the fact that operators are physically safe, whereas their targets are in situ within a conflict. The physical separation may amplify moral dissonance (French, 2010) due to perceptions of what it is to have courage and to be 'at war' from both operators themselves and their peers who may judge them (Holz, 2021). While these problems exist regardless of use of AI, they point to how technologies do not merely change decision-making, they change the decision-making environment with often unintended negative consequences for the operators.

This chapter is especially interested in how individual decision making by the soldier is affected within the larger *jus in bello* processes of war including how the military sustains coherence of effort, organisational capacity and maintain legitimacy (*Box 1*). I argue that moral injury can occur from a perception of being let down by the systems of war fighting (see section 'civilian harms') as well as the individual conduct of soldiers in a conflict.

Reference to *jus in bello* principles may not be enough to manage moral risk. Contrary to some public commentary that supposes that remote pilots are less morally engaged (termed 'moral disengagement') because they are physically distant from their targets, many operators do struggle with taking human life regardless of what others might assure is morally permissible. Enemark (2019) points out that military personnel are able to judge themselves by reference to deeply-held beliefs about right and wrong and how the betrayal of those beliefs causes moral injury and post-traumatic stress disorder (Fani et al., 2021; McEwen et al., 2021; Williamson et al., 2018). For example, an air force study in 2019 found 6.15% of remotely piloted aircraft pilots suffer from post-traumatic stress disorder (Phelps & Grossman, 2021). Drone crews have higher incidence of psychiatric symptoms than pilots of traditionally crewed aircraft (Saini et al., 2021). But as Phelps and Grossman (2021) point out, while remote pilots may have higher rates of psychological effects than regular pilots, it is important to acknowledge that most remote pilots do their job and do not suffer either moral injury or PTSD. So, the claims of harms ought to be considered proportionately.

The opportunity of this chapter is to consider what lessons to take from historical precedent to consider the future operating environment. Without care and consideration, some good apples operating AI and autonomous systems may become 'cooked apples', suffering from a range of negative psychosocial harms from frustration, moral disengagement, lost agency all the way to post-traumatic stress disorder and moral injury due to the way that teaming with technologies changes the environment of decision making and the human role within in.

Considering the above, I argue that moral injury, can come in two forms from working with AI, that of either being a moral witness or becoming moral crumple zone.

Bad, mad, and cooked: Moral responsibility for civilian harms in human-AI teams

## Extreme moral witness

An extreme moral witness is an operator who experiences the humanity of their target intensely, feels emotionally connected to their target and finds the intentional act of harming them challenges their moral beliefs. This may be because the oversight functions required for compliance with international humanitarian law and the laws of armed conflict expose operators to be more situationally aware of the humanity they are harming or putting at risk of harms (Phelps & Grossman, 2021).

A research question is to what degree should operators experience the humanity of their targets? What is the normative ideal towards which designers of human-AI teams ought to aspire? Considering an Aristotelian approach (Aristotle, 350 B.C.E.)—a virtuous human-AI team will create optimal cognitive and affective conditions to ensure human agency and accountability but not put humans through unnecessary suffering or trauma in the course of doing their job. The vices include moral disengagement on the one hand and experience of moral injury on the other hand (see *Table 1*.). Moral disengagement generates a lack of a sense of responsibility whereas moral injury creates a disproportionate sense of moral responsibility, where individuals take on more responsibility for their actions than is perhaps reasonable or required given their role as a part of complex targeting systems, command structures and national imperatives to act that they are obliged to follow. Note extrinsic and intrinsic influences can produce virtuous and vice behaviours. That is to say, virtuous behaviours can be amplified and made more likely within cognitively aligned decision-making environments and be diminished in misaligned cases.

*Table 1* Virtue and vices of moral engagement

| Vice | Virtue | Vice |
|---|---|---|
| Moral disengagement | Moral engagement | Moral injury |
| *Lack sense of responsibility* | *Feels responsible* | *Overdeveloped sense of responsibility* |

Again, there will be sceptics, say absolute pacifists, who wish that soldiers directly and intensely experience any harms they do to others. I argue that soldiers ought to be morally engaged in their decisions, but not morally injured (to the extent possible). That is, there is a difference between what is fair for a murderer to experience (as a consequence of their acts) versus a soldier compelled to harm in accordance with their professional duty. That a nation ought to avoid subjecting service personnel to elevated risk of trauma, depression, anxiety and suicide (Jamieson et al., 2021; Kaldas et al., 2023).

Note, the main argument in this chapter, however, does not require a firm determination with regards to how much soldiers should experience the harms they cause others—I leave this normative framework to each nation and each military to decide. Instead, I argue that whatever the national ethical standards are for engagement, then militaries owe it to their personnel to align their human-AI interface to that standard. Not thinking about a standard and just building AI systems without sensitivity to their effect on human operators is negligent. Determining that operators ought to experience the humanity of their targets to a greater or lesser extent and then making technologies align with this, is responsible conduct.

Bad, mad, and cooked: Moral responsibility for civilian harms in human-AI teams

In terms of how AI might help militaries achieve an ethical standard of engagement, I draw on the social media content moderation literature (Gillespie, 2020). Some theorists support using AI to classify and remove inappropriate content on social media before human moderators experience it. AI could also be used in a similar way in the battlespace, such as evaluating the extent of damage in a battle damage assessment, limiting how much disturbing content human operators must observe. Another technique, called visibility moderation (rather than content moderation) algorithmically alters the content prioritised to viewers of social media (Zeng & Kaye, 2022). A similar approach using AI could limit the amount of disturbing information fed to operators by number of seconds or severity of the content. AI could also dynamically distribute content across a team of human operators so that collectively, the entire battle space and actions within it, were appropriately (i.e. not disproportionately) witnessed, and that the experience was shared by the team. Sharing narratives between human teammates is another way of human experience that is known to help manage post-traumatic stress recovery. Thus, AI can be used to ensure moral engagement rather than moral injury.

## Moral crumple zone

The term 'moral crumple zone' was created to explain cases where human operators are blamed for errors or accidents that are not entirely in their control (Elish, 2019). Moral injury from being a moral crumple zone occurs because an operator shoulders the blame for an incident that has occurred rather than society acknowledging the complex and systemic factors that can lead to adverse outcomes—where the human operator is a component part, and potentially blameworthy, but could not have caused the harm without the technological apparatus within which they made decisions. This concept is not new and has been explored since the 1980s (Dekker, 2013; Reason, 2016; Woods et al., 2017) but gains new nuance with the rise of AI taking over human decision making at higher cognitive levels in ever more complex systems. Operators who fall into a moral crumple zone may have situational awareness, but lack a sense of agency over their situation, that they were a cog in the machine. Or they may lack situational awareness. Perhaps this is because they had to decide too quickly in order to achieve high priority mission objectives, they were slow to intervene on a system that they had previously trusted to operate with little need for oversight, or they had the wrong information upon which to base their decisions. A military example is punishments to personnel in the Médecins Sans Frontières Kunduz hospital incident[2] with no responsibility taken at higher levels for the human-technical systems in place that lead to the tragedy (Donati, 2021)[3]. In the future, two risk situations need to be considered: (1) military responsibility for errors may fail to be attributed due to operational complexity (Bouchet-Saulnier & Whittal, 2018) and/or (2) responsibility is unfairly apportioned to manage political fallout. In each case operators might feel unfairly treated by society who may pick out the human as a poor decision maker, negligent or malfeasant after an incident of civilian harm.

---

[2] "Twelve of the sixteen personnel involved in the bombing of the hospital had been punished with removal from command, letters of reprimand, formal counselling, and extensive retraining. The list included a general officer, the AC-130 gunship aircrew, and the US Special Forces team on the ground", Donati, 2021, p.234.

[3] Medicines San Frontières "asked to know who in the chain of command was ultimately responsible for the forty-two people killed in the hospital that night. That question, along with all the others, was never answered" Donati, 2021, p.235.

Bad, mad, and cooked: Moral responsibility for civilian harms in human-AI teams

So far in this chapter I have defined the key concepts, considered responsibility requirements and the advantages and risks to civilian and military personnel from the introduction of AI into systems. I will now move to new methods to measure human-AI decision making and conditions needed for the attribution of responsibility.

## Human Factors

New methods are needed to explore the degree to which both operators and those who may judge their behaviour will consider them responsible for decisions made with AI and the factors that affect these attributions. In both the moral witness and moral crumple zone cases, modifications of human factors research methods engaged in advance of deployment of human-AI teams can help identify responsibility risks including moral disengagement and moral injury as well as identify optimal decision environments to encourage moral engagement and responsibility for decisions made. These new methods are recommended to build on existing best practice human factors methods both to ensure alignment as well as increase the usefulness of new methods within existing test and evaluation paradigms in Defence contexts of use.

One way to extend is taking existing human factors *naturalistic decision making* (Klein et al., 1993; Zsambok & Klein, 1997) methodologies of cognitive task analysis and add additional probes to critical decision method (CDM) (Table 4.5 CDM probes, Stanton et al., 2017) relevant to evaluating moral responsibility.

Drawing on philosophical analysis (Talbert, 2016) moral responsibility requires:

1. free will (autonomy and agency over one's decisions),
2. situational awareness (knowledge of the circumstances), and
3. capable action (capacity to act during events involving ethical risk in accordance with intent).

I'll start with measuring situational awareness because a long tradition already exists within human factors research to measure it such as SART, SAGAT & SPAM (Salvendy & Karwowski, 2021). Humans working with AI need to be engaged to be effective (Endsley, 2023) and to be accountable for decisions made with AI. Yet, with increasingly complex human-AI systems, operators may not simply lose situational awareness (SA) but lose understanding of the systems themselves including functions and parameters of their operation. This is problematic because humans have the responsibility of overseeing the performance of the AI and ultimate responsibility for decisions made with AI. Humans need to have SA over all aspects of the task including what aspects the AI is undertaking and how well they are working. Humans working with AI systems need to prepared for its perceptual limitations; hidden biases; limits of causal models to predict evolving situations and the extent of the AI's brittleness operating in new situations (Endsley, 2023).

Humans need to be aware of the SA of their human and AI teammates to ensure that team responsibilities are upheld. Humans need to be situationally aware in order to achieve their work responsibilities individually and in teams including perceiving their situation, comprehending it and projecting from the current situation to inform potential future situations. In order to achieve human-AI team SA, the functioning of the AI must be suitably transparent and its actions explainable to the human (Endsley, 2023).

Bad, mad, and cooked: Moral responsibility for civilian harms in human-AI teams

A morally disengaged operator might achieve level 1 SA, perceiving their situation, but not level 2 SA where level 2 requires interpreting what the data means relevant to the goals and decision requirements of the individual, or level 3 where the ramifications for moral disengagement might affect future situations.

There are choices 'cooked' operators may make to try and relieve their experiences such as over trusting automatic functions (possibly relinquishing a sense of responsibility) or deliberating circumventing AI functions to try and gain more control and situational awareness.

Once situational awareness is measured, this leaves measures of free will and capable action to be determined. Critical Decision Method is an established protocol that allows researchers to better understand human decision making within sociotechnical systems (Table 4.5 CDM probes, Stanton et al., 2017).

### Critical Decision Method Responsibility Probes (CDM-R)

Responsibility probes (CDM-R) are described in *Table 2.* under headings of: 'decision agency', 'decision capability' and 'decision responsibility' and are recommended to be used in conjunction with existing situational awareness tools to measure conditions for moral responsibility.

*Table 2.* Decision responsibility probes for Critical Decision Method (CDM-R)

| Decision responsibility | Describe your sense of responsibility for the decision. |
| --- | --- |
| | At what decision points did you feel responsible? At what decision points did you not feel responsible? |
| | How did you determine who or what system was responsible at various decision points? |
| | If there was a decision point at which you felt particularly responsible, describe what aspects of the process affected this feeling |
| | If there was a decision point at which you did not feel responsible? Describe precisely when you no longer felt responsible. |
| | What features affected your sense of decision responsibility? |
| Decision Agency & Autonomy | Describe your sense of agency and autonomy over the decision. |
| | At what decision points did you feel free to apply your own considerations? At what decision points did you not feel free to apply your own considerations? |
| | If there was a decision point at which you felt a strong sense of agency or autonomy, describe what aspects of the process affected this feeling |

Bad, mad, and cooked: Moral responsibility for civilian harms in human-AI teams

|  | If there was a decision point at which you did not feel agency or autonomy? Describe precisely when you no longer felt free to apply your own considerations. |
|---|---|
|  | What features affected your sense of agency and autonomy? |
| Decision Capability | Describe your sense of capability to make the decision. Include skills, knowledge and abilities. |
|  | At what decision points did you feel capable? At what decision points did you not feel capable? |
|  | If there was a decision point at which you felt particularly capable, describe what aspects of this point affected this feeling |
|  | If there was a decision point at which you did not feel capable? Describe precisely when you no longer felt capable. |
|  | What features affected your sense of capability to make the decision? |

Answers to CDM-R prompts will explicate how and when operators feel agency and autonomy over decisions, when they felt capable to act and how informed they were of the situation and technology they were using. Through use of CDM-R human factors researchers would identify points in the decision-making process where operators are more or less likely to experience a sense of responsibility for their actions and may predict how responsibility attributions would play out in post-incident reviews for both legal and ethical determinations. Prompted answers will provide information about gaps in training, skills or knowledge. CDM-R answers combined SA results will provide system and interaction designers areas for improvement of the UX/UI.

More research needs to be done on the experience of responsibility for operators in human-AI teams as well as what factors affect how external adjudicators will judge the degree to which they have acted responsibly and bear responsibility for decisions. It may be that a human operator feels more responsibility for decisions when they under-trust automation and less responsibility when they defer to the automated functions. However, a sense of responsibility may vary throughout different decision points as well as exist at different observation perspectives, transcend a specific decision.

Future research may identify clusters of psychological experiences working with AI systems to allow them to predict the likely effects on operators during decision making. The next section provides information on an AI workplace health and safety framework to identify risks to operators.

## AI Workplace Health and Safety Framework

Human safety issues need to be considered when AI is designed ahead of deployment. Otherwise, AI risks being introduced based primarily on its professional significance rather than its human impact. This section will describe a risk assessment tool towards human safety and AI systems developed by SafeWork NSW (an Australian government regulator) to create an AI Workplace Health and Safety (WHS) score card (Cebulla et al., 2022; Centre for

Bad, mad, and cooked: Moral responsibility for civilian harms in human-AI teams

Work Health and Safety, 2021). To create the tool, researchers used qualitative and quantitative methods including literature review and consultations with AI experts, WHS professionals, regulators and policymakers, representatives from organisations adopting or having adopted AI, and others with knowledge in the field.

Stakeholders agreed that AI influences workflows, both automating tedious and repetitive tasks and creating a new intensity of work, creating new hazards. AI may augment work tasks, offering personnel new methods to improve the quality of their work, which in turn may affect the way tasks are assigned by management, e.g. AI will be used by managers to optimise workflows. The report findings suggested that while a range of psychological, physical, and social risks are associated with the introduction of AI in the workplace, it was in the realm of the psychological that AI was perceived to likely have the greatest impact[4]. Cognitive hazards include information processing, complexity and duration of tasks. Social factors reducing safety were noted including that if AI takes over traditional managerial tasks, it may reduce interactions between workers and managers that could have WHS implications. The effect on social interactions and communication between personnel and their chain of command is relevant for Defence forces looking to implement AI systems (King, 2006). The report noted 'little evidence was found of organisations taking strategic approaches to anticipate the impacts of AI on workplaces beyond the intended process or product change." (p.2).

With regards to the method used relevant to achieving responsible use of AI in military systems, the initial draft of the score card drew on two frameworks, (i) *Australia's AI Ethics Framework* (Department of Industry Innovation and Science, 2019; Devitt & Copeland, 2023; Reid et al., 2023) and (ii) the *AI Canvas* (Agrawal et al., 2018). A simplified framework was developed with three broad categories (human condition, work safety, oversight) and three higher-level steps (ideation, development, application)—see *Table 3*.

*Table 3:* Risk Domains Aggregate eight Australian AI Ethics Principles, Appendix G: AI WHS Protocol (Centre for Work Health and Safety, 2021)

| Human Condition | Worker Safety | Oversight |
|---|---|---|
| Human, social and environmental wellbeing | Privacy protection and security | Transparency and explainability |
| Human-centred values | Reliability and safety | Contestability |
| Fairness | | Accountability |

This new framework was then imbued with principles of good work design considering physical, cognitive, biomechanical and psychological characteristics of a task (Safe Work Australia, 2020, p. 9). The authors note that AI risks may not be visible, detectable physical risks and points of hazards (Cebulla et al., 2022, p. 923). AI is more likely to produce psycho-social risks resulting from AI's dehumanising application. Psycho-social risks involve subjective assessments and are situation-specific, making them harder to measure (Jespersen, Hasle, et al., 2016; Jespersen, Hohnen, et al., 2016). Highly demanding jobs

---
[4] However, a physical risk was noted if workers felt obliged to work faster due to increased surveillance and monitoring.

where employees have little control are likely to increase strain. Whereas giving employees more control of the work (such as the timing, sequencing, speed) reduces strain. This is at odds with the anticipated accelerated decision-making environments envisioned in mosaic warfare (Clark et al., 2020; Devitt, 2023; Hall & Scielzo, 2022)—increasing the likelihood of psychosocial risks to personnel working alongside AI. To increase wellbeing, Human-AI teams ought to prioritise human autonomy, build competence and confidence to be effective with job tasks and to feel connected with people involved with on-the-job tasks (Calvo et al., 2020). Humans find their work meaningful when they have autonomy (Cebulla et al., 2022) and dignity (Bal, 2017) at work. Factors affecting a sense of dignity at work means equality, contribution, openness, and responsibility. Key to this chapter is the link between psycho-social health and a sense of responsibility. Dignity at work signifies work that is meaningful with a degree of responsible autonomy and recognised social esteem. Militaries introducing AI need to consider the wellbeing of their personnel offering purpose and avoiding demeaning, arbitrary authority, unhealthy or unsafe conditions or physical or mental degradation (Autor et al., 2020). The researchers produced the score card with risk rating and an AI WHS Protocol as a guideline for its use.

The Scorecard is extensive (Centre for Work Health and Safety, 2021, pp. 59-71 Appendix F. AI WHS Scorecard) Examples of Risks from AI WHS Scorecard (version 2.0) relevant to military applications are in in *Table 4*.

*Table 4.* Examples of AI ethics risks relevant to military applications from NSW AI WHS Safety Scorecard (Centre for Work Health and Safety, 2021)

| Ethics Risks to WHS | Examples | Potential Military Context |
|---|---|---|
| **Risk of overconfidence in or overreliance on AI system resulting in loss of diminished due diligence** | After a six-month trial of new AI product without incident preventative safety measures are no longer prioritised | An autonomous logistics robot is introduced in a warehouse without consistent safety training and skilling<br><br>An AI targeting tool offers confident and reliable determinations of lawful combatants within a specific context of operations. |
| **Risk of AI being used out of scope** | A productivity assessment tool designed to improve workflow efficiency is used for penalising or firing people | AI tools to measure human performance are unfairly used to recommend promotion, postings and awards including medals<br><br>An AI surveillance tool designed to identify the presence of transient civilians is used to target all |

Bad, mad, and cooked: Moral responsibility for civilian harms in human-AI teams

| | | |
|---|---|---|
| | | humans moving within a target area. |
| **Risk of AI system undermining human capabilities** | AI system automates processes, assigning workers to undertake remaining tasks resulting in progressive de-skilling | Pilots no longer accrue sufficient manual flying hours to respond quickly and competently when an incident occurs.<br><br>An integrated targeting AI using data fusion from multiple sources degrades human reasoning with regards to the likelihood that objects are lawful combatants |
| **Risk of (in)sufficient consideration given to interconnectivity/interoperability of AI systems** | Multiple data sources need integrating, each quality assessed and assured | Network-centric or mosaic warfare fails to coordinate across multiple AI Assets and value systems across allied forces. |
| **Risk of no offline systems or processes in place to test and review veracity of AI predictions/decisions** | An AI tool is used to triage incoming calls to an organisation but the tool provides incomplete answers unable to resolve the query; dissatisfied client complains. | An AI targeting system takes incoming intelligence and produces recommended orders through an LLM. Personnel have no mechanism to review or verify the quality of the recommendations before sending them through their chain of command. |

The introduction of AI into military contexts of use has the potential to reduce repetitive and mundane jobs. Through increasing AI, military personnel may experience greater surveillance and loss of privacy in their roles, which may have less of an impact than if the same conditions were applied in a civilian context. This is because militaries generally have greater leeway on how service persons are expected to behave and be treated including the degree to which their actions are scrutinised. Militaries ought to pay close attention to effects on the war fighter through the introduction of AI such as declines in productivity, efficiency, morale and cohesion. If personnel become disengaged or even hostile within AI-enabled operational circumstances they may threaten the achievement of military and socio-political objectives in the first instance and, if risks not managed, may threaten the reputation of the military organisation. Personnel may even sue the military for the working conditions they experience while being deployed (Fairgrieve, 2014). Up to and including dying in preventable training exercises and routine non-combat incidents (Mohamed, 2021). Thus, while WHS considerations may be different in military contexts than civilian ones from a legal perspective, the future of AI within the military workspace has a myriad of potential harmful effects on personnel in both combat and non-combat circumstances.

And while workplace health and safety (WHS) has limited application in conflicts, there is a developing line of British case law that refers to decisions made in peacetime which result in adverse safety consequences in conflict.

> It will be easier to find that the duty of care has been breached where the failure can be attributed to decisions about training or equipment that were taken before deployment, when there was time to assess the risks to life that had to be planned for, than it will be where they are attributable to what was taking place in theatre. The more constrained he is by decisions that have already been taken for reasons of policy at a high level of command beforehand or by the effects of contact with the enemy, the more difficult it will be to find that the decision-taker in theatre was at fault (Smith and others (FC) v The Ministry of Defence, Supreme Court UK, 2013).

If so, then best practise WHS frameworks might be drawn on in development, acquisition and training ahead of deployment of systems in conflicts to predict and mitigate risks of human-AI military teams. A WHS Scorecard for AI systems (modified for military use) could be used by militaries to estimate effects on operators from working side-by-side with AI such as feeling disempowered, demotivated, exhausted, a sense of a reduction in status or value, anxiety, boredom etc… as well as a sense of reduced or enhanced responsibility.

## Discussion

The future risks a more distributed and fluid like attack by militaries seeking to confuse their enemies and adapt in theatre. This type of command and control, sometimes called 'Mosaic warfare' involves switching control across multiple platforms in an algorithmically optimised operation (Clark et al., 2020). Rather than a war of attrition, the 'decision-centric' mosaic warfare achieves two advantages, it imposes multiple dilemmas on an enemy to prevent it from achieving its objectives and speeds up decision processes. The risk to human connection to these decisions is obvious, if decision making across AI-enabled autonomous platforms is sped up beyond human cognitive capacities, then loss of situational awareness is inevitable. From this there is a loss of knowledge and competence required for moral responsibility for decisions made. Militaries need to consider the human factors of AI enabled mosaic warfare. Extensive simulation, training and exercises with assets ahead of a conflict must occur so that commanders understand the left and right of arc of decisions made in theatre (Devitt, 2023). Human operators need to feel connected to the technology stack they are responsible for. They need to feel their intention expressed through the behaviours of these systems, even if the way the systems achieve this intent is too complex for individual humans to grasp.

## Conclusion

Militaries are responsible for the decision to go to war and the conduct of war itself. The moral blameworthiness of decisions of individuals in conflicts must be considered within the broader decision environments within which they act. AI has the potential to reduce civilian harm and protect soldiers from undue levels of suffering leading to moral injury. However, processes must be instituted to measure changes to decision making in human-AI teams. This is particularly acute with the systemic replacement of cognitive functions by AI that humans would normally be tasked with in targeting decisions to be morally and legally responsible for those decisions. The thesis of this chapter is that good apples can become 'cooked' if they are poorly paired with AI systems including detaching from responsibility for decisions, becoming moral crumple zones, suffering moral injury or becoming extreme moral witnesses. Therefore, militaries must carefully scrutinise AI systems that take away human agency to ensure that safe, meaningful, satisfying, and engaging work is found for humans responsible for decision making and working alongside AI. Combining tools from human factors and workplace health and safety will enable militaries to track, measure and take responsibility for ensuring personnel remain 'good' and not 'cooked'.

Bad, mad, and cooked: Moral responsibility for civilian harms in human-AI teams

Bad, mad, and cooked: Moral responsibility for civilian harms in human-AI teams

Bad, mad, and cooked: Moral responsibility for civilian harms in human-AI teams

Bad, mad, and cooked: Moral responsibility for civilian harms in human-AI teams

Bad, mad, and cooked: Moral responsibility for civilian harms in human-AI teams

Bad, mad, and cooked: Moral responsibility for civilian harms in human-AI teams